\documentclass[aps,twocolumn]{revtex4-1}
\usepackage{amssymb}
\usepackage[export]{adjustbox}
\usepackage{graphicx}
\usepackage{hyperref}
\usepackage[usenames, dvipsnames]{color}
\usepackage[table]{xcolor}
\begin{document}
\author{Shikha Bhadoria}\email{shikha.bhadoria@physics.gu.se}
\author{Thomas Blackburn}\email{tom.blackburn@physics.gu.se}
\author{Arkady Gonoskov}\email{arkady.gonoskov@physics.gu.se}
\author{Mattias Marklund}\email{mattias.marklund@physics.gu.se}
\affiliation{Department of Physics, University of Gothenburg, Sweden}

\title{Mapping the power-law decay of high-harmonic spectra from laser-plasma interactions}

\begin{abstract}
Visible or near infra-red light can be manipulated to produce bursts of coherent extreme ultraviolet (XUV) or X-rays via the relativistic high-order harmonic generation process when a laser irradiates a solid plasma target.
The intensity of the spectral components of the reflected signal decays with increase of harmonic order and the efficiency of this non-linear process largely hinges on how prompt this decay is.
This is governed by the conditions of the laser-plasma interaction for which various models have been proposed. At relativistic intensities, a spectrum exhibiting a power-law decay with an exponent of $8/3$ or $4/3$ is often stated.
Here, we analyse the dependence of this exponent on interaction parameters including the angle of incidence, the carrier envelope phase, intensity of the laser and the pre-plasma length, and discuss opportunities for optimization.
Our simulations show that, rather than there being one universal exponent, the spectral decay is a continuous function of the laser-plasma interaction parameters.
\end{abstract}
\maketitle

\section{Introduction}\label{Sec-intro}

When a relativistically intense laser is focused on a plasma target, high-order harmonics of the incident laser light are generated around the specular direction. 
High-order harmonic generation (HHG) by the laser-plasma interaction process can serve as a compact source of coherent XUV or soft X-ray pulses of attosecond duration. 
Unlike the HHG mechanisms in gaseous media \cite{Krausz2009}, HHG mechanisms driven by laser-plasma interactions can withstand much higher field intensities and thereby offer an advantageous alternative\cite{Teubner2009}.
The generated bright ultrashort pulses from HHG process have multiple applications ranging from studying warm-dense matter\cite{Dobosz2005}, non-destructive inspection techniques\cite{Brenner_2015}, 
studies of strong-field quantum electrodynamics \cite{gonoskov.arxiv.2021} and capturing the ultra-fast physical, chemical or biological processes by serving as the fastest cameras\cite{Krausz2009}.

Many experiments have demonstrated HHG from plasmas\cite{rodel.prl.2012,Kahaly2013,Dromey2006,Dromey2012,Teubner2009,Kormin2018,dollar.prl.2013} and distinct theoretical models explain the non-linear process responsible for the observed spectra\cite{PGibbon1996,CWE2006Quere,Baeva2006,anderBruegge2010,AG2011,Gordienko2004,Lichters1996,Pirozhkov2006}. The characteristics of the emission strongly depend on the initial parameters and configuration of laser-plasma interaction. Two forces, the laser driver force ($\propto a_0 $ laser pulse amplitude in relativistic units) and plasma restoring force ($\propto n_e$, electronic density normalised by the classical critical density $n_c$), compete with each other giving rise to a convenient similarity parameter $S=n_e/a_0$\cite{Gordienko2005}. All mechanisms fundamentally depend on the coupling of laser's energy with that of the electrons at the plasma-vacuum interface and have distinct regions of dominance, efficiencies and signatures of the reflected spectrum. In the case of weakly relativistic laser intensities, $a_0\leqslant 1$, coherent wake emission (CWE), where electrostatic plasma oscillations are excited by the wake of Brunel electrons\cite{CWE2006Quere} and the emission process is limited  by the local plasma frequency, dominates. At higher intensities, the relativistic HHG mechanism is described by `relativistic oscillating mirror' (ROM) model, where the harmonic emission is due to the Doppler upshifted laser frequency due to the relativistic motion of a reflecting oscillating point\cite{Lichters1996,Baeva2006}. This model assumes the Leontovich boundary condition implying that the incoming and outgoing energy fluxes at some oscillating point are equal to each other at each instant of time, i.e. the energy is not accumulated by the plasma.
At higher intensities and when moderate plasma densities are chosen (so as not to allow the onset of relativistic transparency), a significant part of incident laser energy ($\sim 60\%$) can get deposited into plasma, due to the relativistic motion of electron nanobunches against relatively stationary ions. This process leads to release of short bursts of energy (with $>100$ times the initial intensity) and 
  can be modelled by
  `relativistic electron spring' (RES) model \cite{AG2011,AG2018}. 
  The spectral trends of the bursts can be described by
  coherent synchrotron emission (CSE)\cite{anderBruegge2010,Dromey2012} where the accelerated nanobunches generate coherent emission while following synchrotron-like trajectories.
    \begin{figure}
    \centering
    \includegraphics[height=0.25\textheight,width=0.48\textwidth]{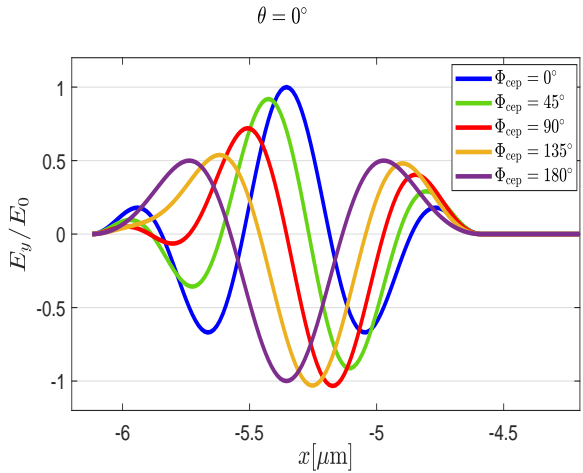}
    \caption{The variation of the peak amplitude of the laser with the carrier envelope phase. }
    \label{fig-cep}
    \end{figure}
The high-order harmonics of the input laser in the reflected emission are phase locked and their spectral intensity decays with increasing frequency. Due to the non-linear nature of interaction between the electrons at the interaction surface of the solid plasma and intense laser driver, the generated harmonics show a long-tail frequency distribution. Theoretical modelling and PIC simulations suggest that this decay follows a power law\cite{Baeva2006,anderBruegge2010,Gordienko2004}.
The ROM model predicts that exponent of power-law decay of harmonic intensity is $8/3$, based on the similarity theory highlighting that this law is universal and is independent of laser-plasma interaction conditions\cite{Baeva2006,Gordienko2004}. However, this is challenged in ultra-relativistic regime where the interaction is too complex and a range between 5/3 to 7/3 is predicted\cite{Boyd2008}. The RES model, on the other hand, predicts a slower exponential decay\cite{AG2011} which can be fitted to a power law with an exponent $~1.31$\cite{AG2018}. The CSE model predicts a value in the range 4/3 to 6/5\cite{anderBruegge2010}. A recent study on these exponents and cutoffs (that mark a faster decay) shows that this can even exceed 8/3 for a wide range of parameters\cite{Edwards2020}. The exponent is important because it determines how quick the decay is, which affects the IR to XUV/X-ray conversion efficiency and so the feasibility of HHG as a radiation source.
  
In this paper, we investigate the power-law exponent exhibited by the reflected spectrum for a laser-plasma system with a fixed similarity parameter, but different configuration parameters.
These parameters include the laser's angle of incidence $\theta$, the carrier envelope phase $\phi_{\rm CEP}$ and the length scales $L_p$ of an exponentially falling pre-plasma.
These parameters determine the initial conditions of interaction and play a pivotal role in deciding which mechanism will dominate or interfere\cite{Kahaly2013}.
We show that the exponent of spectral power-law decay for $S=64.87$ spans a range of values from $5.4$ to $2.2$, in contrast to specific values that are widely considered as universal (for ROM, $8/3$, or CSE, $4/3$).
We identify an optimal region within this parameter space ($\theta$, $\phi_{\rm CEP}$ and $L_p$), where the high-harmonic generation process is most efficient and has a slower decay, where the RES mechanism is in play.
Subsequently, choosing the best value of $\phi_{\rm CEP}$ from this optimum region, we extend our investigation to higher laser intensities (lowering the $S$ parameter down to 1.9), which further supports our findings.
  
The paper is organised in the following way.
In Sec.~\ref{Sec-simulations}, we describe the details of a series of simulations performed with varying laser-plasma configuration parameters.
Thereafter in Sec.~\ref{Sec-Eref-m} we discuss the characteristics of the reflected electric field as a function of the chosen parameters.
The exponent of the spectral intensity decay and conversion efficiency are presented.
Then we extend our parameter scan to higher intensities in Sec.~\ref{Sec-hi-intens}.
In Sec.~\ref{Sec-summary} we summarize and discuss our findings.
Appendix~\ref{appendix} discusses the numerical reliability of our findings and presents analytical estimates of the optimal pre-plasma scale length.

  \section{Simulation details}\label{Sec-simulations}

  A series of PIC simulations has been performed using ELMIS-1D\cite{AGThesis} where a p-polarised laser of wavelength $\lambda_0=$765 nm and intensity $I_0=$ $8.9\times 10^{19}W/$cm$^2$ (with vector potential $ a_0=e E_0 / m_e \omega_0 c=6.1659 $), irradiates a $Si^{4+}$ plasma at an angle of incidence $\theta$. Here $E_0$ is the electric field strength of the laser,  $e, \omega_0$,$m_e$  and $c$ are the elementary charge, the laser frequency, the electron mass and the velocity of the light in vacuum. This oblique-incidence problem is reduced to one dimension by the Bourdier technique \cite{Bourdier1983}, where a frame boosted by $c \sin\theta$ along the plasma surface is chosen. The laser has the following profile: $E_y = E_0' \Theta( -\psi) \Theta(-4\pi - \psi)\sin (\psi/4) [0.5 \cos(\psi/4) \sin(\psi) + \sin(\psi/4) \cos(\psi+\phi_{\rm{CEP}})]$. Here $ E_0'=E_0\cos\theta$ is the wave amplitude in the moving frame, $\Theta$ is the Heaviside step function, $\phi_{\rm{CEP}}$ is the carrier envelope phase, $L_{m}$ is the maximum length in laser's propagating direction $x$, $\psi$ is the phase coordinate $2\pi (x -ct)/\lambda' $, with $ \lambda' =\lambda_0/\cos\theta$ being the wavelength of the laser in the moving frame. This configuration is close to a Gaussian profile of duration $\sim$3 fs at full-width-half-max, FWHM and has been chosen to ensure that the vector potential of a few-cycle laser has the same value as the coordinate approaches infinity. A strong dependence of the peak electric field with the $\phi_{\rm{CEP}}$ of the ultra short pulse laser can be observed in  Fig.~\ref{fig-cep}.
  The plasma is initialised with a temperature of 1 keV and the ions in the plasma are mobile with masses $ m_i/m_e=28\times1836,$ ($m_i$ being the mass of the ion). The plasma slab (with a step function profile) is 2 $\mu$m wide and has a peak density of 400 $n_c$, where $n_c$ is the classical critical density for a laser pulse ($n_c = m_e \omega_0^2 / 4 \pi e^2$) leading to $S=64.87$. This is preceded by a preplasma that decays exponentially from the peak with a scale length $L_p$. 
   \begin{figure}
  \centering
  \includegraphics[height=0.23\textheight,width=0.42\textwidth]{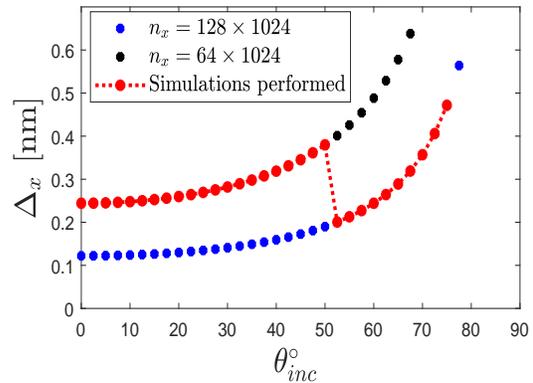}
  \caption{Dependence of spatial resolution on the angle of incidence in moving frame with Bourdier technique. }
  \label{fig-theta-cellsize}
  \end{figure}
  These simulations have been performed varying the angle of incidence $\theta \in [0^\circ, 75^\circ]$ with a step size of $2.5^\circ$, the carrier envelope phase $\phi_{\rm CEP} \in [0^\circ, 180^\circ]$ with a step size of $6^\circ$ and the pre-plasma scale length is varied from  $L_p \in [0.08,0.73]\mu$m with a step size of $0.05\mu$m. This amounts a total of 14415 simulations. The length of the simulation box is $L_x = 16\lambda'$ extending from $-8\lambda '$ to $8\lambda'$. Fig.~\ref{fig-theta-cellsize} shows the variation of the the spatial cell size ($\Delta_x = L_x/n_x$, with $n_x$ being the number of cells) with the angle of incidence of laser. It can be clearly seen that with increasing angle of laser incidence, the cell size increases very rapidly in the boosted frame. Thus, these automated simulations have been performed in two sets with different resolutions and are limited to a $\theta_{max}=75^\circ$ so as to trade off numerical accuracy against run-time. The number of cells in these simulations for angle of incidence from $[0^\circ, 50^\circ]$ is $64\times 1024=65536$ and in those from $[50^\circ, 75^\circ]$ is $128 \times 1024=131072$. This has been done in order to ensure that the spatial resolution $\Delta_x$ remains small (between 0.24 nm for $0^\circ $to 0.47 nm for $75^\circ$) even at larger angles to avoid numerical instabilities, as the size of the simulation box in the transformed frame is $\theta$-dependent. The number of particles per cell is 85 and there are 256 timesteps per plasma period.
   
  \section{The reflected spectrum}\label{Sec-Eref-m}
    \subsection{Peak electric field}
    \begin{figure*}
    \centering
    \adjincludegraphics[height=0.71\textheight,width=1\textwidth,trim={2cm 0 0 0},clip]{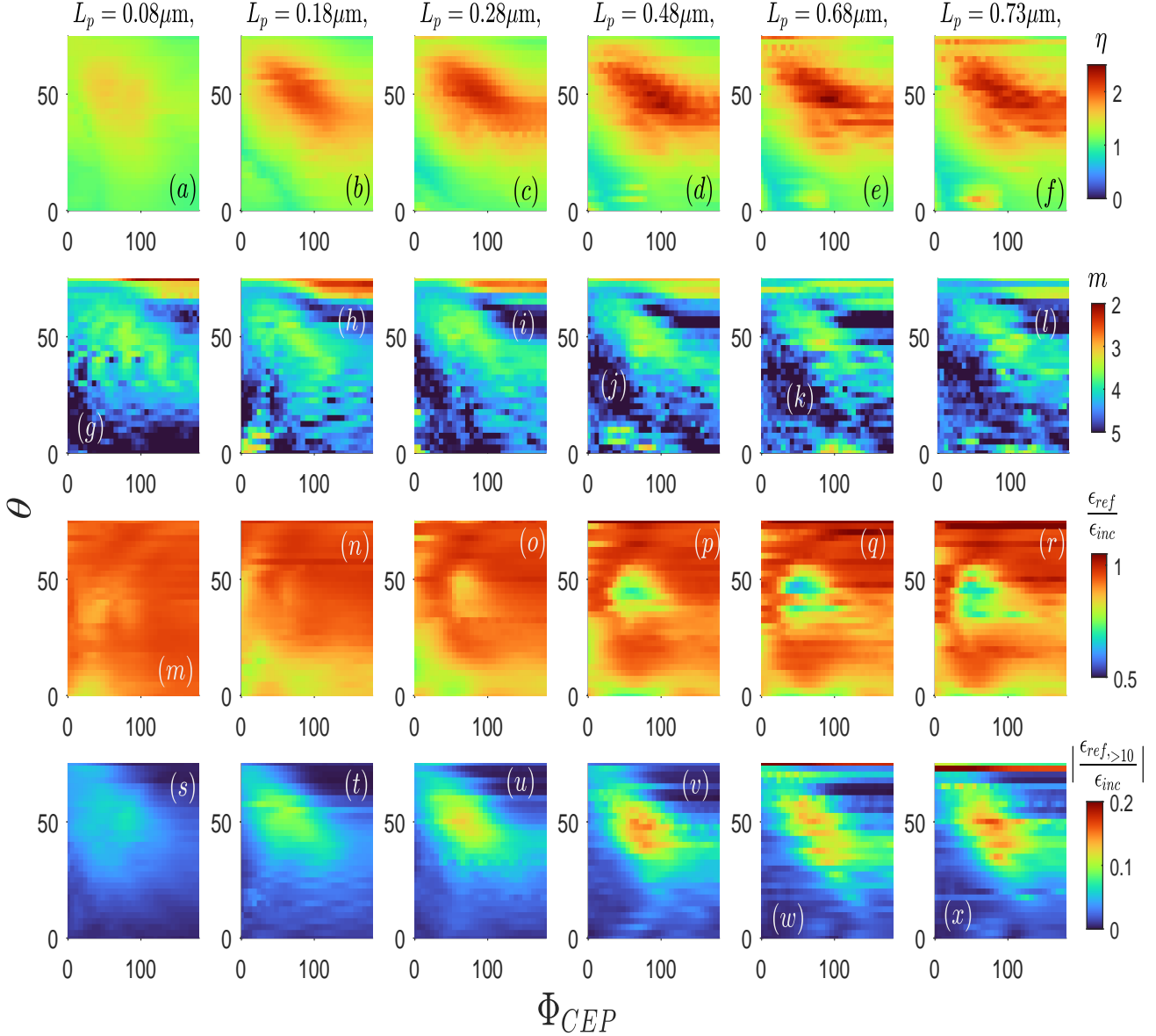}
    \caption{The exponents of a decaying power-law exhibited by the re-emitted spectrum for each simulation.}
    \label{fig-Eref-m}
    \end{figure*} 
  The top row of Fig.~\ref{fig-Eref-m} shows the the highest factor by which the incident laser field has been amplified in the re-emitted spectrum, based on these simulations. Here, each subplot (panels $[(a)-(f)]$) shows the maximum strength of the reflected electric field, normalised by the incident laser field amplitude ($\eta=|E_{\rm{ref}}/E_0|_{\rm{max}}$), as a function of all angles of incidence ($\theta \in [0^\circ, 75^\circ]$) and carrier-envelope phase ($\phi_{\rm CEP} \in [0^\circ, 180^\circ]$) for certain pre-plasma scale lengths (labelled on the top of each panel).
  The pre-plasma scale length is increased from left to right with $L_p \in [0.08,0.73]\mu$m in the each row (corresponding to $[0.105,0.954]\lambda_0$).
  Interestingly in each subplot, one can distinctly observe the formation of an optimum region in $\theta-\phi_{\rm CEP}$ space, where the field amplification is significantly larger ($\eta>>1$) than in the other regions. Beginning with the first subplot of shortest pre-plasma length ($L_p \in 0.08 \mu$m), where the plasma is closest to a perfectly sharp mirror-like boundary, this region is relatively small, $ 40^\circ \leqslant \theta \leqslant 60^\circ $ and $ 30^\circ \leqslant \phi_{\rm CEP} \leqslant 100^\circ $.  Moreover, in panel Fig.~\ref{fig-Eref-m}(a) $\eta$ is not remarkably above unity in most of the region, and reaches 1.8 in the optimum region. The color gradient around this region indicates a smooth increase from the lower $\theta$ and $\phi_{\rm CEP}$ values followed by a similar subsequent decay. In low $L_p$ simulations, ROM-like conditions dominate the laser-plasma interaction and as the $L_p$ is gradually increased, the optimum region expands in $\theta-\phi_{\rm CEP}$ space (notably wider in $\phi_{\rm CEP}$) yet approaches a much narrower region of maximum field strength gain ($\eta \sim 3$ times).
  This augmentation, related to pre-plasma length, is expected as longer pre-plasma lengths would lower the effective $S$ parameter, allowing an improved energy accumulation in plasma favouring the RES mechanism\cite{TGB2018,AG2011}. One can see in (panel $(e)$) that a very narrow peak of $\eta $ within this optimum region at $[\theta,\phi]= [55^\circ,90^\circ]$ which starts showing clearly from $L_p=0.28\mu$m and reaches its maximum at $L_p=0.68\mu$m, after which it disappears. 
  
\begin{figure}
  \centering
  \includegraphics[height=0.21\textheight,width=0.48\textwidth]{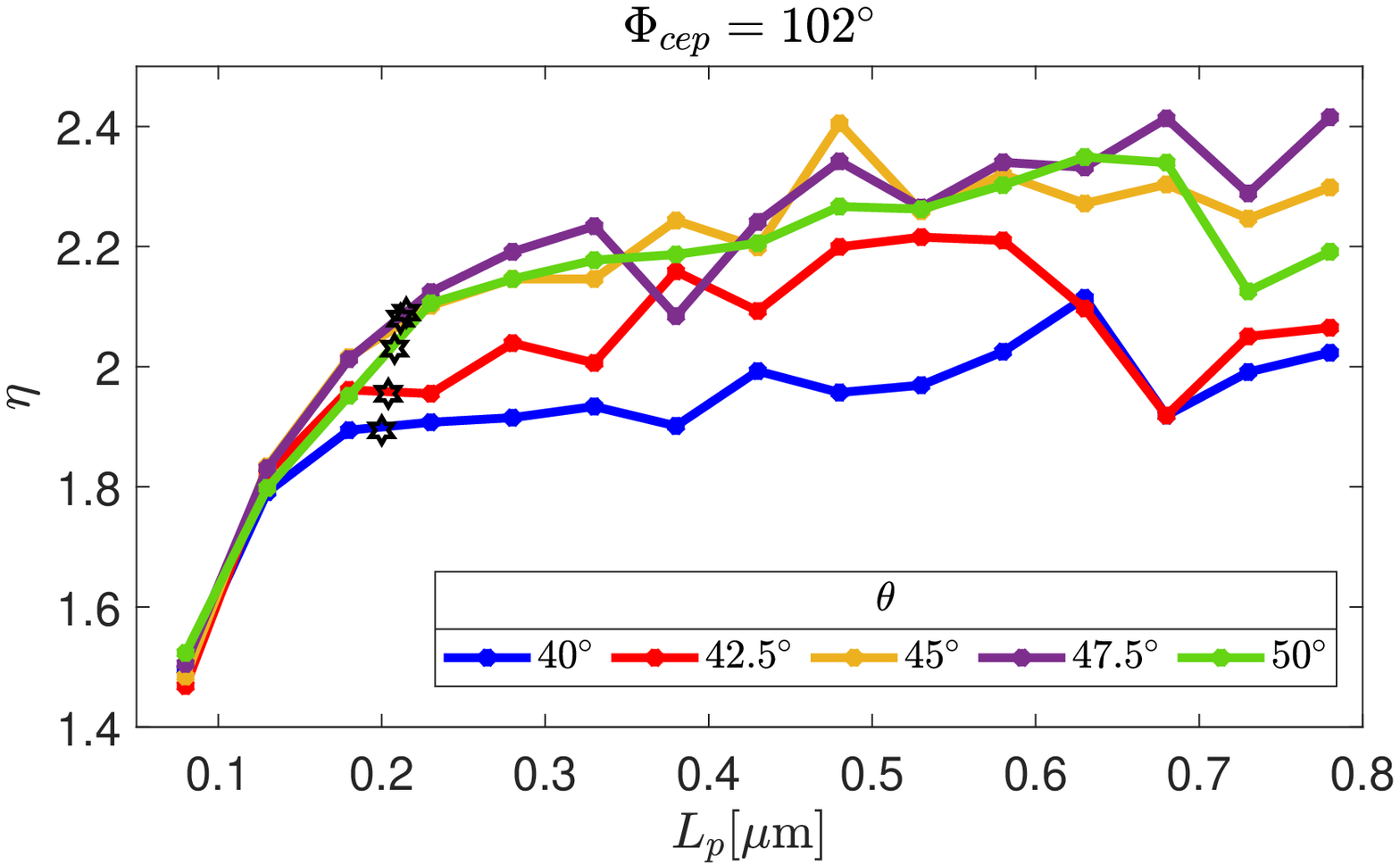}
  \caption{The dependence of normalised peak reflected field with the pre-plasma lengths. The black hexagrams mark the optimal length scale as predicted by eqn.\ref{eqn-Lopt}. }
  \label{fig-Lp}
  \end{figure}

  This can be better seen in Fig.~\ref{fig-Lp}, which shows the variation of $\eta$ with different pre-plasma lengths for a range of angle of incidence $\theta\in [40^\circ,50^\circ]$ and a particular value of $\phi_{\rm CEP}=102^\circ$ from the optimum region. Here we see a sharp increase in $\eta$ as the pre-plasma length is increased up to $\sim 0.2\mu$m beyond which $\eta$ grows very slowly. 
  A black star on each line of Fig.~\ref{fig-Lp} marks a theoretical prediction of the optimal pre-plasma scale length, as given in Sec.~\ref{appendix:PreplasmaDynamics}.
  \begin{equation}\label{eqn-Lopt}
      L_{opt}\sim \frac{\lambda_0 (1+\sin\theta)}{2\pi},
  \end{equation}
   which ensure to $S_{\rm eff}\sim 1$. 
   The length scales below $L_{opt}$ indicate the interaction scenarios when $S_{\rm eff}>1$. 
For an ultra-short laser pulse, like the one used in this work, $\phi_{\rm CEP}$ has a significant impact on the amplitude and position of the first peak of the electric-field profile, which drives the initial stages of the laser-electron interaction. The effectiveness of the electron nanobunching subsequently affects the efficiency of the high-harmonic emission. Simulations indicate that $\phi_{\rm CEP}\sim [90^\circ-100^\circ] $ allow for a favourable interaction scenario for the case considered.

  \begin{figure}
  \centering
  \includegraphics[height=0.18\textheight,width=0.48\textwidth]{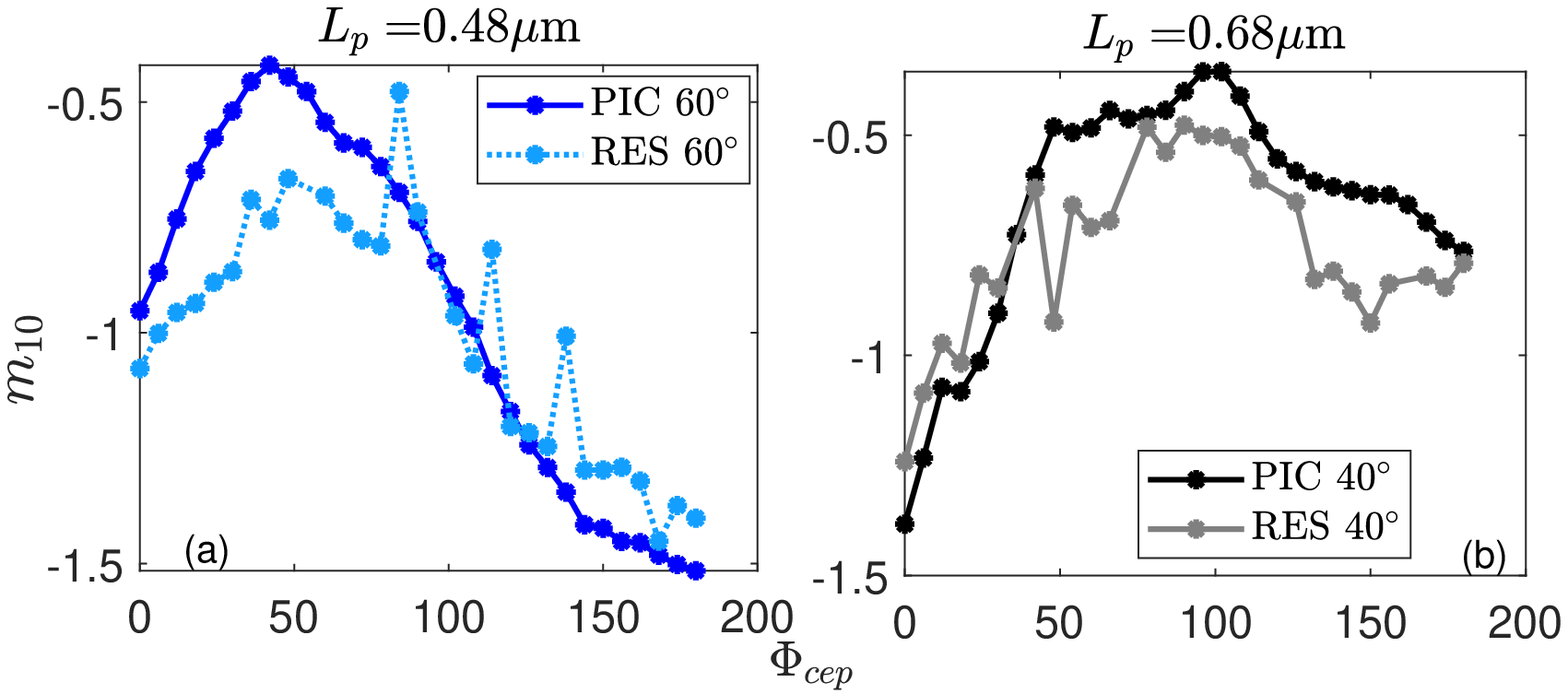}
  \caption{The $\phi_{cep}$ dependence of the slope of first 10 harmonics from PIC simulations and the RES model\cite{AG2011}. 
  }
  \label{fig-phicep}
  \end{figure}
  
  There is an appearance of a small well-defined island (Fig.~\ref{fig-Eref-m}(f)) of potentially enhanced emission at lower angles (at about $[\theta,\phi]= [5^\circ,70^\circ]$) for $L_p > 0.4$, where the field amplification is much enhanced ($\eta=2$ for longer pre-plasma length $L_p=0.73\mu$m). Also, at very high $\theta>70^\circ$ at $L_p=0.73\mu$m, there is yet another increase in the value of $\eta \sim 2$. However, these could also be an outcome of numerical effects due to lower resolution at large $\theta$. In order to convincingly identify this, we plot the peak field amplitude within the plasma region after the interaction as a measure of numerical effects in Fig.~\ref{fig-gof} (Appendix.\ref{appendix}). The impact of resolution at high-angle of incidence leading to rising numerical instability is captured in Fig.~\ref{fig-gof}, in the top row (panels $[(a-f)]$), where each subplot shows the maximum value of laser-normalised field in the spatial region of plasma ($E_{\rm{err}}$), far away from the reflected signal whose peak is shown in Fig.~\ref{fig-Eref-m}. In Fig.~\ref{fig-gof}, one can clearly see that the error increases with $\theta$ and is close to zero at lower values. This clarifies that the increased value of $\eta$ in the top layer of Fig.~\ref{fig-Eref-m} at $\theta>70^\circ$ at $L_p=0.73\mu$m is potentially an outcome of numerical effects. Nevertheless, the distinct broad optimal region for enhanced emission as well as the small island (at about $[\theta,\phi]= [5^\circ,70^\circ]$) may be relied upon.
    \subsection{Decay exponent}
  The spectrum obtained after reflection of the laser in each configuration is unique as its intensity falls with increasing laser harmonic order.
  A slower decay of intensity or a smaller exponent is naturally favourable for effectively up-converting the laser frequency. In order to relate the optimum regions, in the first row of Fig.~\ref{fig-Eref-m}, where the reflected electric field boost is maximised, we plot the exponent of the decaying spectral power-law in the second row of Fig.~\ref{fig-Eref-m} (panels$(g)-(l)$). This exponent is calculated by finding the slope of a linear fit to the log-scaled intensity $\ln I(\ln \omega)$ spectrum obtained from PIC simulations.
  It should be pointed out that the spectrum obtained from these simulations excludes the field contribution due to numerical effects shown in the top row of Fig.~\ref{fig-gof} shown in App.\ref{appendix}. The corresponding adjusted $R^2$ value to the fit on each spectrum is plotted in the lower row of Fig.~\ref{fig-gof} indicating the goodness of fit captured in the exponent of the power law in Fig.~\ref{fig-Eref-m}. Clearly, the adjusted $R^2$ value is above 0.9 in most of the cases (meaning $ 90\%$ of the spectral data can be explained by this fit) until about $\theta=65^\circ$, above which it varies from 0.8 to 0.7 with $\phi_{\rm CEP} $. It should be noted that the range of harmonic spectrum chosen for data fitting with $\theta<65^\circ$ is $2<\omega/\omega_0 < 100$ (corresponding to the energy being 2.62 eV to 162 eV, $\hbar \omega_0 = 1.62$eV). For $\theta>65^\circ$ a bit longer range is taken, $2<\omega/\omega_0 < 250$ as this leads to an overall improvement of the fit ($R^2_{\rm{min}}$ from 0.6 to 0.7) on the spectral data implying a subtle deviation from power-law behaviour. Having clarified the regions of a good fit on decaying spectral power-law, the exponent trends in the second row of Fig.~\ref{fig-Eref-m} (panels $(g)-(l)$) can be analysed more meaningfully.

  In panels $(g)-(l)$ of Fig.~\ref{fig-Eref-m}, it can be seen that there is a smooth dependence of the exponent of the power-law on the experimental parameters. The clear global observation is the convergence of the exponents to a relatively lower value in the same $\theta-\phi_{\rm{CEP}}$ space as the optimum region of high-$\eta$ in top row of the same Fig.~\ref{fig-Eref-m}. This indicates the region of slower decay of harmonics with the potential to find rare events of high-frequency-high-intensity radiation. Interestingly, the range of exponents widely deviates from 8/3, the value expected by the universal law even in cases with smaller pre-plasma lengths, where ROM-like conditions dominate (see $L_p= 0.08,0.13$~$\mu$m). In the Fig.~\ref{fig-Eref-m}~$(g)$ showing $L_p= 0.08$~$\mu$m, the exponent in the optimum region converges to a moderate value of 3.5. Nevertheless, this optimum region is widely spread out in $\theta-\phi_{\rm{CEP}}$. In the non-optimal regions in Fig.~\ref{fig-Eref-m} $(g)$, the exponent value can also reach up to 5.8 (at around $\theta=0,\phi_{\rm{CEP}}=156^\circ$) implying an undesired fast decay of harmonics and a significant deviation from the universal value even where the plasma configuration would allow ROM like conditions. As the pre-plasma lengths are increased towards the right from $L_p= 0.13$~$\mu$m, the RES mechanism of HHG begins to dominate as the effective $S$ parameter of interaction begins to get closer to unity. Here, the optimum region starts to converge to a more specific region of a lower absolute value of exponent. In Fig.~\ref{fig-Eref-m}-$(k)$ ($ L_p=0.68$~$\mu$m), the value of exponent reaches upto 3.1 for a very specific $\theta=50 ^\circ$ and $\phi_{\rm{CEP}}=80^\circ$. This precisely coincides with the point of maximum $\eta$ in the optimum region in top row of Fig.~\ref{fig-Eref-m}.

In order to crosscheck our simulation results, we also numerically solve the RES equations~\cite{AG2018} for a subset of the parameter scans under consideration.
The dependence of the decay exponent on $\phi_{cep}$, observed in the PIC simulations, is consistent with that predicted by the RES theory, which can be seen in Fig.~\ref{fig-phicep}.
Here we restrict the fit to the frequency range corresponding to the first ten harmonics, which yields a decay exponent $m_{10}$.
This is because the frequency range in which the RES model is accurate is itself intensity-dependent~\cite{AG2018}.
  
  In other regions of Fig.~\ref{fig-Eref-m}-$(k)$, the furthest exponent value reaches up to 5.4. Moreover, at high $\theta-$high $\phi_{\rm{CEP}}$  regions (particularly Fig.~\ref{fig-Eref-m}-$(h,i)$) the value of exponent for the spectral power law reaches around 2.2. However, the corresponding adjusted $R^2$ value (Fig.~\ref{fig-gof}-$(h,i)$ in App.\ref{appendix}) reduces sharply to 0.68 in this region from 0.95 in the optimum region, which indicates a clear deviation from power-law behaviour. Interestingly, the small island of enhanced emission at lower angles (at about $[\theta,\phi]= [5^\circ,70^\circ]$) for $L_p > 0.4$ also has a lower exponent ($m=3.1$ for $L_p=0.68\mu$m, same as in the optimum region) implying the possibility of stable high harmonics at near-normal incidence. Overall for all simulations with a fixed $S$-parameter, the exponent of spectral power-law decay is found to vary in the range $5.8>m>2.2$. The lower limit of the values, that mostly appears at low-$\theta$-low-$\phi$ and high-$\theta$-high-$\phi$ regions, is very close to the ones found in a recent study that reports the exponent value to be in the range $6>m>4$ for a single cycle laser with $\theta=30^\circ, S\sim 60, L_p=0,\phi_{\rm CEP}=0$\cite{Edwards2020}. The narrowing down of the optimum region of stable harmonics with increasing $L_p$ can be related to HHG generation by the RES mechanism.
  
\subsection{Conversion efficiency}

  The third row of Fig.~\ref{fig-Eref-m} [panels $(m-r)$)] shows the fraction of laser energy ($\epsilon_{inc}$) going into the entire reflected spectrum ($\epsilon_{ref}$). It can clearly be seen that at lower pre-plasma lengths (esp. $L_p=0.08\mu$m, panel $(m)$ ) almost all of the laser energy goes into the reflected signal. This is understandable as this configuration with a relatively sharp boundary allows for ROM-like conditions. As the pre-plasma lengths are increased, one can observe that in the optimum region of enhanced and stable re-emission, the laser energy in the re-emitted signal reduces in a small area which precisely coincides with the optimum regions in the corresponding panels above. This points to the region of RES dominance, where this reduction in laser energy going into the re-emitted spectrum can only be because of its deposition into plasma-electron nanobunches.  Clearly, even at lower pre-plasma lengths $L_p=0.08\mu$m (Fig.~\ref{fig-Eref-m}$(m)$), there is a certain amount of laser energy that is being deposited into plasma ($\sim 10-12\%$). Later, a clear trend of lowering of the laser energy imparted into the re-emitted spectrum in the optimum region can be seen in Fig.~\ref{fig-Eref-m}$(o-r)$. In Fig.~\ref{fig-Eref-m}$(q)$, at $L_p=0.68\mu$m, this total laser energy being imparted into the reflected signal is reduced to 0.62 at the same location (of $\theta-\phi_{\rm{CEP}}$) where we observe the maximum field amplification and the lowest exponent of power law ($\theta=50 ^\circ$ and $\phi_{\rm{CEP}}=90^\circ$) in the corresponding panels above Fig.~\ref{fig-Eref-m}$(a-l)$. This would imply that about $38\%$ of the incident energy could be used up in electron heating at the solid boundary. 
  Nevertheless, the fraction of incident energy ($\epsilon_{inc}$) going into higher order harmonics is increased. This is captured in the lowest row Fig.~\ref{fig-Eref-m}[panels $(s-x)$)] which clearly shows that the energy imparted into the reflected spectrum above the 10th harmonic ($\epsilon_{ref,\geq10}$) also increases precisely in the same optimum region. Here as well, one can observe the expected trend of RES enhancement of energy going into higher harmonics at longer pre-plasma lengths within the same $\theta-\phi_{\rm{CEP}}$ region. So, longer pre-plasma lengths that modify the effective $S$- parameter are favourable in compressing laser-plasma energy into higher harmonics as also seen in Ref.~\cite{TGB2018}.

  \section{Higher laser intensities}\label{Sec-hi-intens}
     \begin{figure*}
  \centering
  \adjincludegraphics[height=0.71\textheight,width=1\textwidth,trim={2cm 0 0 0},clip]{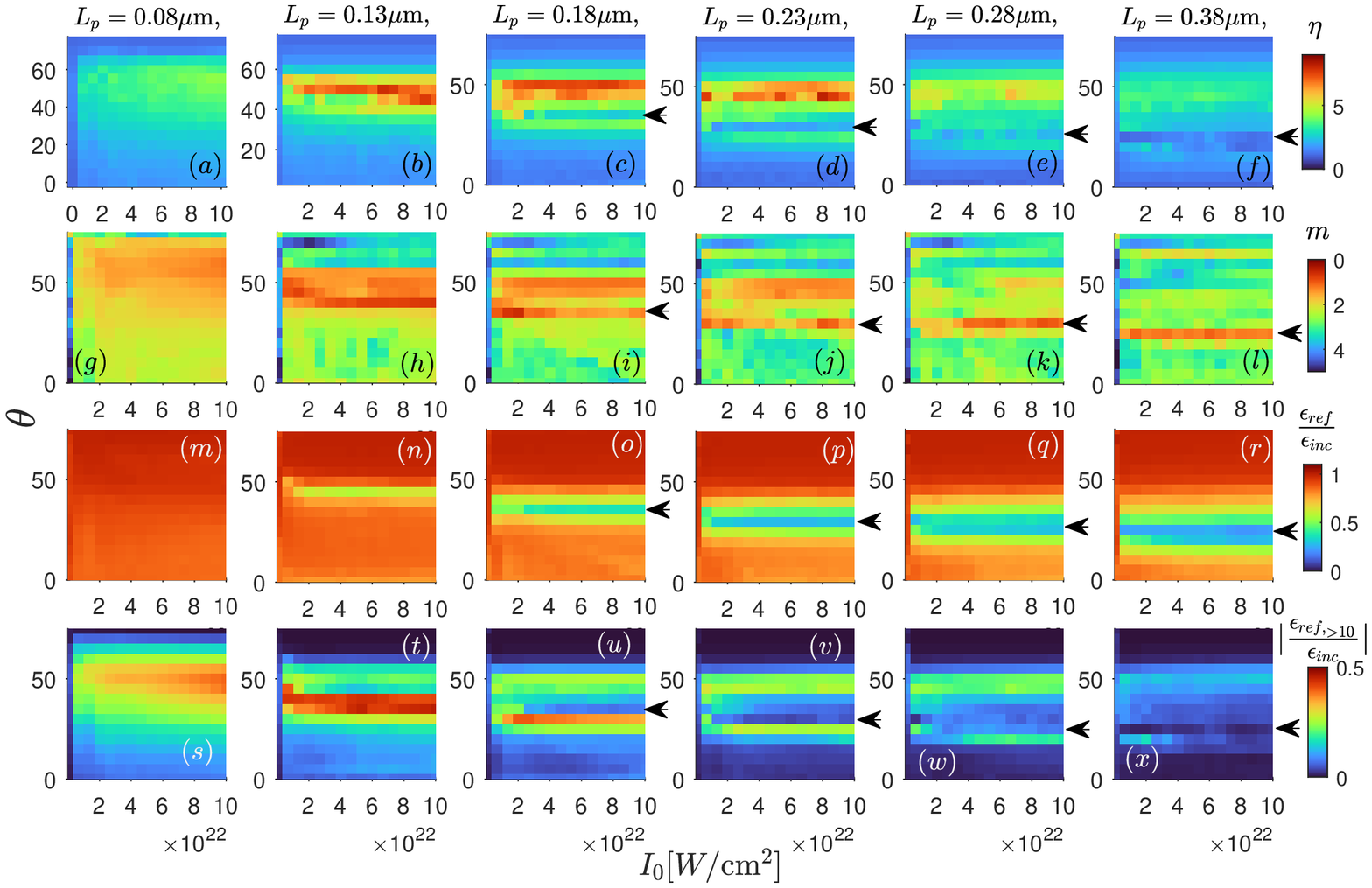}
  \caption{The exponents of a decaying power law exhibited by the emitted spectrum for each simulation. Arrows are explained in the text.}
  \label{fig-In1}
  \end{figure*}
  
  Fixing the carrier-envelope phase $\phi_{\rm CEP} = 100^\circ$ from the optimum region, we now investigate the dependence of the exponent of spectral power-law decay for different intensities $I_{0} \in [8.9 \times 10^{21} , 1.0089 \times 10^{23}]$ W/cm$^2$. The other parameters are same as before, leading to a decreased bulk $S$ parameter. This is shown in Fig.~\ref{fig-In1} which shows the peak electric field boost $\eta$ (top row), power-law exponents (second row), and the fraction of laser energy ($\epsilon_{inc}$) into the entire reflected spectrum ($\epsilon_{ref}$) and spectra above the 10th harmonic ($\epsilon_{ref,\geq10}$) (lower two rows respectively). This has been arranged in the same fashion as Fig.~\ref{fig-Eref-m}. One may notice small arrows systematically pointed at certain angles for all panels above $0.13\mu$m. These arrows point to those precise angles for each pre-plasma length where the physical quantities in each row display an 
  unusual trend (seen as a sharp change in colour) and shall be discussed later in this section.
  \begin{figure}
  \centering
  \includegraphics[height=0.21\textheight,width=0.48\textwidth]{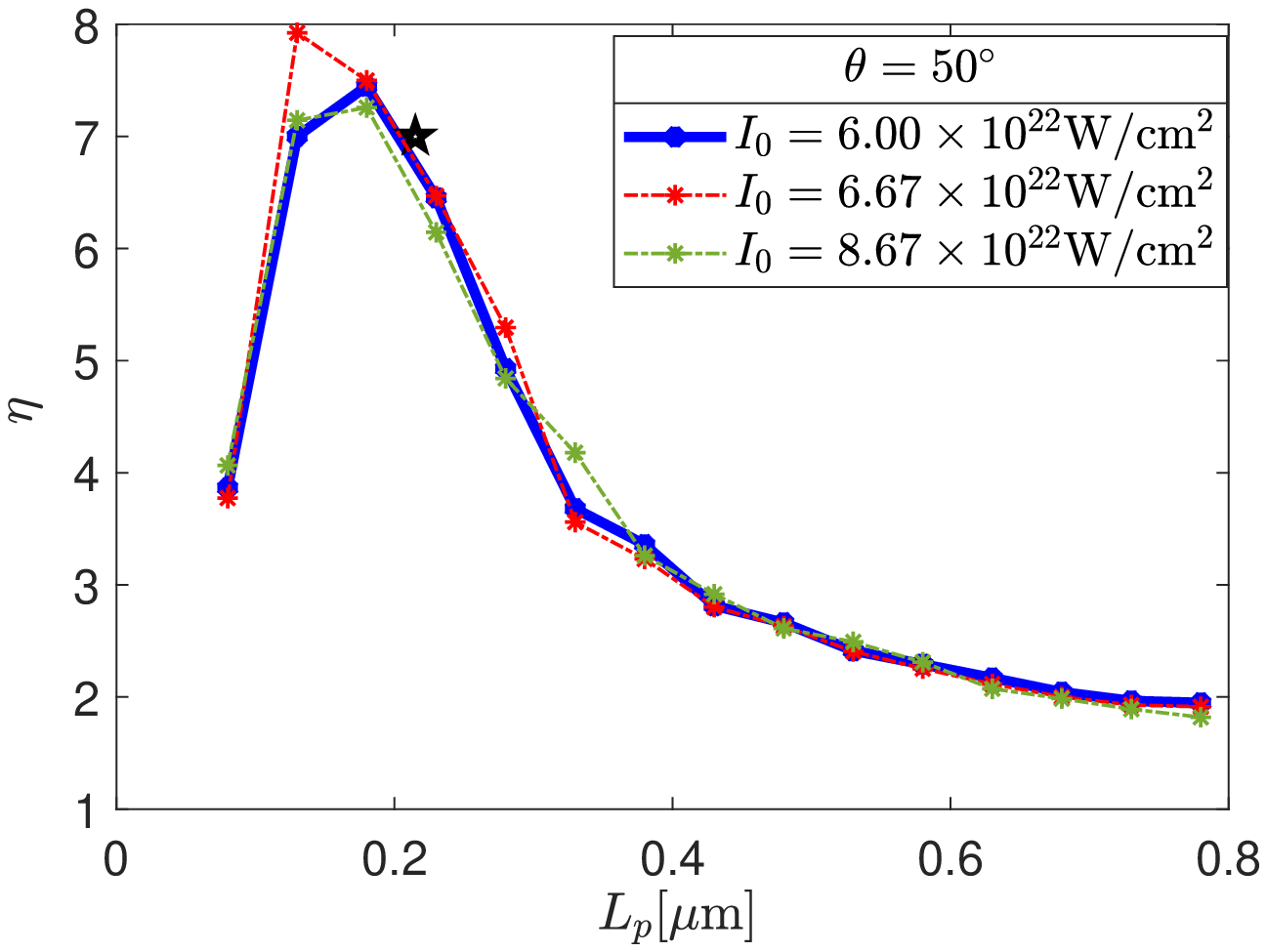}
  \caption{The dependence of normalised peak reflected field with the pre-plasma lengths. The black hexagrams mark the optimal length scale by eqn.\ref{eqn-Lopt}. }
  \label{fig-Lp-i0}
  \end{figure}
  \begin{figure}
  \centering
  \includegraphics[height=0.3\textheight,width=0.45\textwidth]{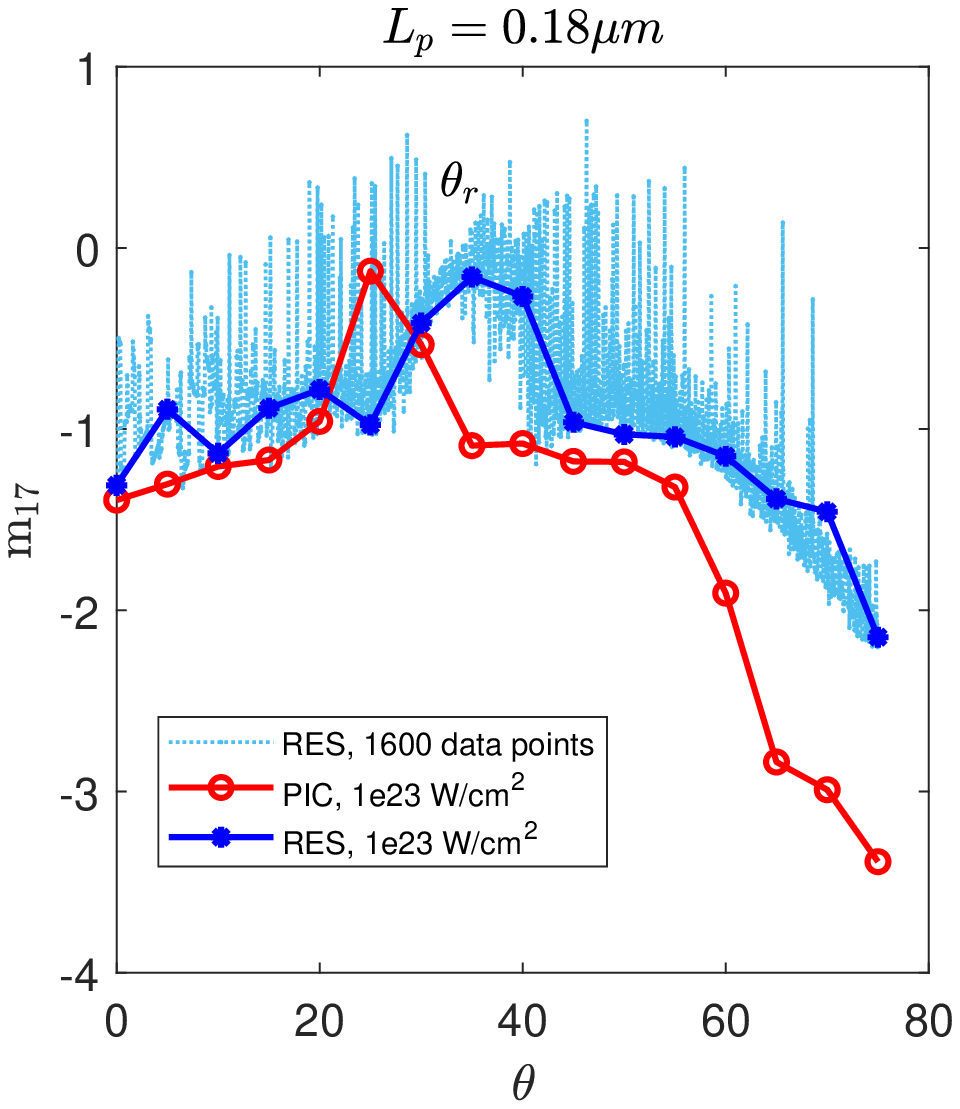}
  \caption{The $\theta$ dependence of the slope of first 17 harmonics from PIC simulations and RES model. The unusual behaviour at $\theta_r$ predicted by both RES and PIC }
  \label{fig-resonance}
  \end{figure}
 
  First we see in the top row of Fig.~\ref{fig-In1} that, like the previous section, there is a clear optimum around $50^\circ$ where $\eta$ is higher. In Fig.~\ref{fig-In1}(a), we see a larger $\eta\sim4$ than before that is also $\theta-$dependent. This is expected, as the relativistic motion of electrons in the plasma enhances RES. The electric-field increase $\eta$ quickly increases with $L_p$ and peaks around $L_p=0.13-0.23\mu$m (Fig.~\ref{fig-In1}$(b-d)$). Unlike in Fig.~\ref{fig-Eref-m}, this optimal region of high $\eta$ decreases rapidly with increasing $L_p$ after reaching its peak (see Fig.~\ref{fig-In1}$(e-f)$), as seen in Ref.~\cite{TGB2018}. This can largely be due to the fact that the laser intensity is held constant in the previous set of simulations, whereas here the laser intensities are varied over orders of magnitude.
   Around $L_p=0.18\mu$m (Fig.~\ref{fig-In1}(c)), $\eta$ can be as high as 8. This pre-plasma length of maximum gain is also closer to the optimal length predicted by RES theoretical estimates. This can be seen in Fig.~\ref{fig-Lp-i0}, which shows a distinct peak. The black star in the figure is the theoretical estimate of $L_{opt}$ computed as before by Eq.~(\ref{eqn-Lopt}). This length is also consistent with the widely reported finding that high-harmonic generation is optimized for a pre-plasma with a similar length~\cite{Kahaly2013,rodel.prl.2012,dollar.prl.2013}. 
  
  The second row shows that, even for higher intensities, the decay exponent does not appear to have a universal value.
  At the optimum angle of 50$^\circ$ and $L_p=0.13-0.18$, the value of the exponent reaches as much as 1.16.
  Furthermore, similar to the row above, the exponent at this angle reduces with $L_p$ to around 2.5.
  This points to the efficient conversion region that can be seen in the two rows below. The conversion efficiency is reduced to around $\sim 0.8$ in this optimal region (50$^\circ$ and $L_p=0.13-0.18$). In the same region of lowest panel (Fig.~\ref{fig-In1}(t,u)), the energy in higher harmonics is higher, pointing to XUV conversion efficiencies of around 20$\%$.
  
  Beyond $L_p=0.13$ (Fig.~\ref{fig-In1}(o-r)), the loss in the reflected spectrum energy (up to 0.2) at lower angles is not directly related to but is an offshoot of RES mechanism. The arrows mark certain lower angles for each length scale, that break up the plasma surface. At these angles (with $\phi_{cep}=100^\circ$), a part of electron nano-bunches that drive RES are extracted from the surface and co-propagate with the re-emitted radiation. The radiation therefore loses some of its energy to electrons and the spectrum deviates from a power law.
  This can be seen consistently in all the panels.
  This lower angle $\theta_{r}$ points to some $L_p$-dependent resonance, and can be distinguished as the abrupt line of lower values of all parameters under discussion [$\eta$, $m$, $\epsilon_{ref}$ and $\epsilon_{ref>10}$]. These have also been corroborated with higher resolution simulations and are also in agreement with RES theory which can be seen in Fig.~\ref{fig-resonance}. Here, the RES theory and PIC simulations predict a similar trend and a resonant peak in the exponent values is clear (This seems shifted by few degrees potentially due to sparse PIC data). Excluding these odd lines at $\theta_{r}$, one may clearly see the distinct stable radiation signatures at 50$^\circ$ and $L_p=0.13-0.18$ with an exponent going beyond $4/3$~\cite{anderBruegge2010} or $1.31$~\cite{AG2011}.
  \section{Summary and Conclusions}\label{Sec-summary}
  To summarize, we have carried out a systematic study of the decay exponent of the spectral intensity of high harmonics, for a fixed laser and plasma (same bulk S parameter), as well as higher intensities (lowering bulk S parameter), for different configuration parameters, including the angle of incidence and the carrier envelope phase of the laser and the pre-plasma length of the plasma. We have shown that the exponent of the power law can span a range of values for different parameters ($5.8$ to $1.1$).
  Even for configurations where plasma mirror reflects the majority of the incident energy, the exponent non-universally deviates from $8/3$.
  Optimal parameter spaces of enhanced emission and their dependence on the laser-plasma configuration parameters are identified and related to the mechanisms at play.
  We show that at lower pre-plasma lengths the region of moderate exponent is spread out widely in $\theta-\phi_{\rm{CEP}}$ space while, at longer pre-plasma lengths, a lower exponent value can be reached in a more precise $\theta-\phi_{\rm{CEP}}$ space. We have investigated these exponents in relation to other features of the re-emitted spectrum, such as reflected peak field amplitude and laser-energy conversion efficiency.
  We have highlighted regions in the parameter space that are not optimal for applications, despite the fact the spectral decay is relatively weak, because the conversion efficiency and peak-field strength are not large.
  Thorough understanding of the dependence of the exponent on laser-plasma parameters, to which this study contributes, is paramount to the usefulness of high-harmonic generation as a potentially rich source of XUV or X-rays.
  
  \section{Appendix}\label{appendix}
  
  \subsection{Numerical accuracy}\label{appendix:NumericalAccuracy}
\begin{figure*}
  \centering
  \adjincludegraphics[height=0.32\textheight,width=1\textwidth,trim={2cm 0 0 0},clip]{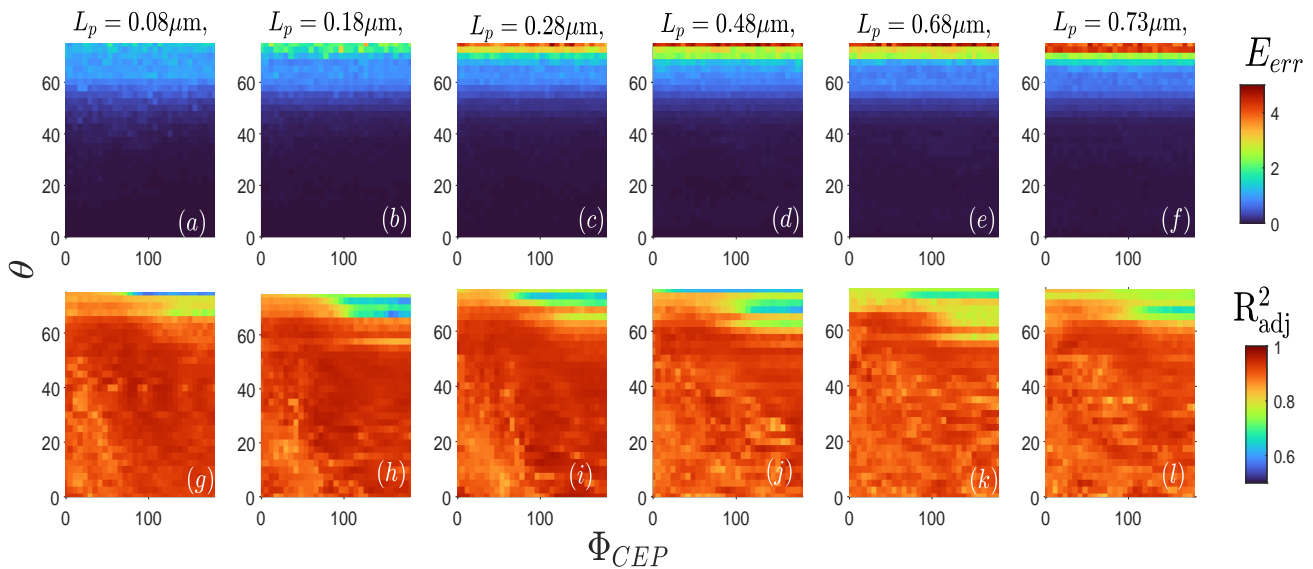}
  \caption{The top row shows the error defined as peak field value within the plasma region after its interaction with the laser (normalised by the peak field of the laser). The bottom row shows the adjusted $R^2$ value indicating the goodness of fit for the exponents of the spectral power-law scaling for high-order harmonics (shown in Fig.~\ref{fig-Eref-m}).}
  \label{fig-gof}
  \end{figure*}

  Simulations are always subject to possible numerical inaccuracy. In order to gauge their reliability, we plot the peak field amplitude within the plasma region after the interaction as a function of the same parameter space in the top row of Fig.~\ref{fig-gof}. This serves as a measure of numerical error and clearly, the results in the optimal region are devoid of the numerical instabilities in the plasma region. We also present the goodness of fit of the slopes (shown in the lower row of Fig.~\ref{fig-Eref-m} that is found by a linear fit of the simulation spectra) in the lower row of Fig.~\ref{fig-gof}. The adjusted $R^2$ value is close to 0.9 in the majority of the fits performed, while its reduction at high $\theta$ indicates a deviation from power-law behavior. It must be pointed out that the re-emitted spectrum on which the fitting is performed, precludes any involvement of the fields in the plasma region originating as numerical noise (shown in the top row of Fig.~\ref{fig-gof}).  
 
  \subsection{Preplasma dynamics}\label{appendix:PreplasmaDynamics}
 Here we repeat the analysis in Section 3 (\emph{Preplasma}) in \citet{TGB2018}, but for an exponential, rather than linear, density ramp. The incident laser pulse penetrates the preplasma up to the point where the force of the charge separation field, established between the electron layer and the background ions, is balanced against the total magnetic force, established by the laser field and the uncompensated ion current. Assuming that all the electrons in the region $z' < z'_{max}$ are displaced, and that the ions are not displaced, we find that this balance occurs at
	\begin{equation}
	z'_{\rm max} =
		-L_p \ln \!\left[ \frac{2\pi L S}{\lambda (1 + \sin\theta)} \right],
		\end{equation}
	\begin{equation}
	L_p > \frac{\lambda (1 + \sin\theta)}{2 \pi S}.
	\label{eq:PenetrationDepth}
	\end{equation}
The condition on $L_p$ in \ref{eq:PenetrationDepth} ensures that the laser pulse
does not break through the density ramp. This will be the case for all the results
presented here.

This result gives the maximum displacement of the electron layer, which
occurs when the plasma interacts with the peak intensity of the laser pulse.
We can derive the instantaneous displacement of the electrons $\xi(t')$ by assuming that the temporal profile of the laser varies adiabatically and therefore that force balance applies at every instant of time between the
charge separation field and the instantaneous magnetic field of the laser. We therefore replace $a_0$ in \ref{eq:PenetrationDepth}, which is implicit
in the $S$ factor, with $a(\phi) = a_0 \exp[-\phi^2/(2\tau^2)]$, where the phase
$\phi = \omega t' \cos \theta$ and $\tau = \omega T / (2 \sqrt{\ln 2})$. The plasma density encountered by the laser at phasefront $\phi$ is $n_0 \exp[\xi(\phi)/L]$, from the definition of the density ramp. The value of the normalized density at this point follows as $s(\phi) = n_0 \exp[\xi(\phi)/L] / [ a(\phi) n_c]$. The dependence on phase cancels out, leaving $s(\phi)$ equal to a constant
we label the ``effective $S$'' parameter:
	\begin{equation}
	S_{\text{eff}} = \frac{\lambda (1 + \sin\theta)}{2\pi L_p}.
	\label{eq:EffectiveS}
	\end{equation}
This is, rather interestingly, independent of the bulk value of $S$, due to the self-similar properties of the exponential function~\cite{dollar.prl.2013}.
If we increase $n_0$ at fixed $L$, \ref{eq:PenetrationDepth} tells us that the point at which the pulse is reflected, $z'_{\rm max}$, moves further
out into the preplasma. This compensates for the increase in $n_0$, because the local value of the density $n_e(z'_{\rm max}) = n_0 \exp[z'_{\rm max}/L_p]$.

The relativistic electron spring model predicts that XUV generation is optimized at $S \sim 1$, $\theta \sim 60^\circ$. We can identify that the scale length $L$
necessary to reduce \ref{eq:EffectiveS} to unity is eq.\ref{eqn-Lopt}.
This is consistent with the widely-reported finding that high harmonic generation is optimized for a preplasma with $L \simeq 0.25 \lambda$~%
\cite{rodel.prl.2012,Kahaly2013,dollar.prl.2013}. 
  \subsection*{Acknowledgements}
  The research is supported by the Swedish Research Council (Grant No. 2017-05148 and 2020-06768). The authors acknowledge computational resources provided by the Swedish National Infrastructure for Computing (SNIC).
  \bibliography{HHG-expo.bib}

\end{document}